\newcommand{\be}{\begin{equation}}
        \newcommand{\ee}{\end{equation}}
        \newcommand{\ba}{\begin{eqnarray}}
        \newcommand{\ea}{\end{eqnarray}}
        \newcommand{\ban}{\begin{eqnarray*}}
        \newcommand{\ean}{\end{eqnarray*}}

\newcommand{\C}{{\mathbb C}}

\documentclass[11pt]{article}
\usepackage {graphicx,latexsym}
\usepackage{amsfonts}

\begin{document}

\title{
Loop quantization from a lattice gauge theory perspective}
\author{Jos\'e A Zapata
\footnote{e-mail: \ttfamily zapata@math.unam.mx}
Instituto de Matem\'aticas UNAM \\ 
A.P. 61-3, Morelia Mich. 58090, M\'exico
}

\date{August 6, 2004}
\maketitle

\begin{abstract} 
We present an interpretation of loop quantization in the framework of lattice gauge theory. 
Within this context the lack of appropriate notions of effective theories and renormalization group flow exhibit loop quantization as an incomplete framework. 
This interpretation includes a construction of embedded spin foam models 
which does not rely on the choice of any auxiliary structure (e.g. triangulation) 
and has the following straightforward consequences: 
\begin{enumerate}
\item The values of the coupling constants need to be those of an UV-attractive fixed point 
\item The kinematics of canonical loop quantization and embedded spin foam models are compatible 
\item The weights assigned to embedded spin foams are independent of the  
2-polyhedron used to regularize the path integral, $|J|_x = |J|_{x'}$ 
\item An area spectrum with edge contributions proportional to $l_{\rm PL}^2 (j+1 / 2)$ is not compatible with embedded spin foam models and/or canonical loop quantization 
\end{enumerate}
\end{abstract}

It is usually said that the kinematics of canonical loop quantization \cite{LQreviews} 
is well understood and that  the main gap towards a complete understanding of the theory is on the dynamical side (that there  is no certainty about the correctness of the space of solutions to the Hamiltonian constraint). 
The problem of studying the macroscopic behavior of the theory and connecting it with low energy physics is often regarded as the first mandatory test, but it is not involved in any defining aspect  of the theory. 

These ideas sharply contrast with the modern point of view behind other approaches (perturbative or non perturbative) to field theory in which one works with effective theories and the renormalization group is an integral part of the approach. In particular lattice gauge theory's working philosophy, which is governed by Wilson's renormalization group, is in sharp contrast with what is usually assumed in loop quantized theories. 
Needless to say that the origin of this situation is that loop quantization is geared towards treating background independent theories. In that context the meaning of renormalization is unclear and even the concept of scale needs to be adapted. On the other hand, in principle loop quantization should also be able to treat theories that use a background metric. 
Thus, the formalism should be applicable in a context where renormalization is 
a well understood 
necessary process which can be implemented in several alternative frameworks. 

It is also relevant to to remind the reader that the relation between canonical loop quantization and spin foam models \cite{SF} 
is not fully understood. 
Below we describe the minimal requirements that a full theory of spin foam models should have to be compatible with canonical loop quantization. 

Starting from the canonical framework in terms of initial data one would like to see spin foam models provide a map from the kinematical Hilbert space to the 
Hilbert space of physical states. 

Taking the four dimensionality of spacetime as primordial it would be desirable that the TQFT-like structure 
of spin foam models played a central role. Then one would say that a spin foam model is compatible with canonical loop quantization if: (i) the partition function's boundary data induced a Hilbert space which coincided with the kinematical Hilbert space of canonical loop quantization,  and (ii) the space of classes defined by the map induced by $\Sigma \times I$ (the map must be degenerate) coincides with the physical Hilbert space of canonical loop quantization. 

It was argued that the discretization used by spin foam models was a clear obstruction to compatibility. Thus, a natural objective was to find a way to get rid of the discretization of spin foam models. The group field theory approach \cite{GFT} and embedded spin foam models \cite{SFC} are two strategies to accomplish such an objective. 


Here we will present loop quantization within the framework of lattice gauge theory. 
This is the main result of this work and it brings the following two contributions. 
First, loop quantization is seen as part of a working philosophy centered in the renormalization group. However, since an adequate method to implement the renormalization group is not given, loop quantization is exhibited as an incomplete framework. 
Second, the result includes a notion of an embedded spin foam model which is fully compatible with the kinematics of canonical loop quantization%
\footnote{
This interpretation of loop quantization does not introduce any background structure; it is compatible with the representation of the diffeomorphism group defined by loop quantization. 
}. 
It should be acknowledged that (up to now) the only examples of background independent embedded spin foam models have no local degrees of freedom. Thus, the complete success of our proposal rests in the finding of non trivial examples. 


A {\em general lattice with boundary} $\cal X$ has faces, edges and vertices in the bulk (boundaries of faces are composed by edges and boundaries of edges by vertices), 
and it has edges and vertices in the boundary. The boundary edges and vertices 
form a lattice denoted by $\partial \cal X$ which 
hosts the  boundary data --the connection that is kept fixed when calculating the path integral. The lattice is called general because it is not required to have regular connectivity. The space of connections in  $\cal X$ can be trivialized and presented as a product of a copy of the gauge group $G$ per edge, 
and the gauge group is identified with a copy of $G$ per vertex. 
We briefly review this framework in the language of spin networks and spin foams mainly to fix notation; for detailed introductions see \cite{SF}. 

The space of functions of the boundary data can be made into a Hilbert space because the boundary data is essentially a collection of group elements, and the Haar measure in each copy of the group allows us to integrate the needed functions \cite{hamlat} 
\[
{\cal H_{\partial X}} = 
L^2({\cal A_{\partial X}}/ {\cal G_{\partial X}}, d\mu_{\partial {\cal X}})
= \C [SN(\partial \cal X)] . 
\]
In the last equality ${\cal H_{\partial X}}$ is presented as the space of linear combinations of the spin network orthonormal basis with finite norm. A spin network $j \in SN(\partial \cal X)$ is a function of the connection defined by an assignment of irreducible representations to the edges of $\partial \cal X$ and an assignment of intertwiners to its vertices. 

To define the path integral one constructs a weight based on the desired 
lattice regularization of the 
action with real or imaginary time and 
integrates over the connection degrees of freedom%
\footnote{
There may be other degrees of freedom as long as they do not belong to the boundary data. One then has to find a way to integrate them out. 
}
 living in the interior edges \cite{wilson, SF} 
\[
\rho_{ _{\cal X}}(A)=  \int_{{\cal A}_{{\cal X}^{\circ}}} 
d\mu_{\cal X} \hbox{\rm Weight}(A')
= \sum_j \left( \sum_{J | \partial{J}=j}  | J |_{ _{\cal X}}
\right) j(A) , 
\]
where the integral runs over configurations $A'$ satisfying $A'|_{\partial X} = A$. 
In the  last equality, the path integral --a function of the connection-- is written as a linear combination of spin networks, 
$\rho_{ _{\cal X}}(A)=  \sum_j \rho_{ _{\cal X}}(j)  j(A)$. In addition the coefficients 
$\rho_{ _{\cal X}}(j)$ are expanded in terms of spin foams. Each spin foam  $J \in SF(\cal X)$ is 
an assignment of irreducible representations to the faces of $\cal X$ and an assignment of intertwiners to its bulk edges. It is a history (in the spin network representation) compatible with $\partial{J}=j$, 
and $| J |_{ _{\cal X}}$ is its weight. For a detailed introduction see \cite{SF}. 

Note that the covariant lattice kinematics on ${\cal X}$ and the canonical kinematics on $\partial {\cal X}$ are compatible. Moreover, Kogut and Susskind used the transfer matrix to define a canonical lattice theory completely compatible with Wilson's covariant lattice theory. 

\medskip

\begin{description}
\item[{\em Working philosophy and interpretation:}] 
In lattice gauge theory one first specifies the scale at which the study will be performed. 
Then the modeling is made on a lattice $\cal X$ (appropriate for the scale) that hosts an effective theory whose coupling constants take the value $\lambda^i(\cal X)$ specified by the renormalization group flow generated by a renormalization scheme. 

In principle, the effective theory at  $\cal X$ could serve as a seed to construct further effective theories based on ``coarser lattices" 
integrating out the ``unnecessary details" (extra degrees of freedom not relevant at the scale of interest). Thus, 
$\lambda^i = \lambda^i({\cal X})$ are running coupling constants determined by the renormalization group flow. 

At a single  $\cal X$, the coupling constants take definite values $\lambda^i({\cal X})$ which define the effective theory at $\cal X$. 
For this lattice theory we can use the spin network/spin foam language described above. The boundary data of the effective theory lives in ${\cal H_{\partial X}}$, and the effective partition function is $\rho_{ _{\cal X}}$. 

If the renormalization group flow manages to define a theory in the ``limit of the finest lattice," this theory should be treated as a {\em fundamental theory}. The theory in this ``finest lattice" 
has a sublattice structure described below; it has the kinematical structure of a loop quantized theory. 
The fundamental boundary data lives in ${\cal H}_{\partial M}$, and the normalized fundamental partition function is $\rho^n_M$ (to be defined below (\ref{sprescription})). 
\end{description}

\medskip

Let $S({\cal X})$ be the set of {\em sublattices} of lattice ${\cal X}$. This set 
is partially ordered by inclusion and directed in the direction of refinement.
The same structure is present in the set  $S({\cal \partial X})$. 

Consider a set of nested sublattices of $\cal \partial X$, 
$\gamma_1 \leq \gamma_2 \leq \ldots \leq {\cal \partial X}$. 
A connection on a lattice induces a connection on any of its sublattices. Then we have the following natural maps 
${\rm Fun}({\cal A}_{\gamma_1}) \rightarrow {\rm Fun}({\cal A}_{\gamma_2}) \rightarrow 
\ldots \rightarrow {\rm Fun}({\cal A}_{\cal \partial X})$ (denoted by 
$i^*_{\gamma' \gamma}$). 
The measures $d\mu_{\gamma_i}$ are compatible in the sense that 
$
\int_{{\cal A}_{\gamma_1}} d\mu_{\gamma_1} f(A) = 
\int_{{\cal A}_{\gamma_i}} d\mu_{\gamma_i} (i^*_{\gamma_i \gamma_1}f)(A) , 
$
which means that we have a set of nested Hilbert spaces 
\be
\label{nestedH}
{\cal H}_{\gamma_1} \leq {\cal H}_{\gamma_2} \leq 
\ldots \leq {\cal H}_{\cal \partial X}, 
\ee
where we have omitted the isometric inclusions $i^*_{\gamma' \gamma}$.

Functions of the lattice connection $f \in {\rm Fun}({\cal A}_{\gamma_1})$ induce operators acting by multiplication on any ${\cal H}_{\gamma_n}$ with 
$\gamma_1 \leq \gamma_n$. 

Vector fields on ${\cal A}_{\cal \partial X}$ induce derivative operators 
$i_{{\cal \partial X} \gamma_i *}v: 
{\rm Fun}({\cal A}_{\gamma_i}) \to {\rm Fun}({\cal A}_{\gamma_i})$. These operators are compatible in the sense that 
$i_{{\cal \partial X} \gamma' *} \circ i_{\gamma' \gamma *}v = i_{{\cal \partial X} \gamma *}v $, which allows the family of compatible momentum operators on the sublattices of ${\cal \partial X}$ 
to be the building blocks for all the ``momentum" operators on 
${\cal H}_{\cal \partial X}$ \cite{diffopAL}. 

Spin network functions form an orthonormal basis for 
${\cal H}_{\gamma} = \C [SN(\gamma)]$. For our set of nested sublattices we have 
$SN({\gamma_1}) \leq SN({\gamma_2}) \leq  \ldots \leq SN({\cal \partial X})$, 
where to include a spin net from $\gamma_1$ to  $\gamma_2$ one assigns trivial representations and intertwiners to the extra edges and vertices. 
Now let $\gamma(j)$ be the subgraph of ${\cal \partial X}$ composed by the edges to which $j$ assigns nontrivial representations (and their connecting vertices). Then we also have 
\[
{\cal H}_{\gamma} = \C [ \{ j\in SN({\cal \partial X}) | \gamma(j) \leq \gamma \} ] \subset {\cal H}_{\cal \partial X} . 
\]

Spin foams are compatible with the nested structure of the sublattices in a similar form as spin networks are. For a set of nested 
sublattices of $\cal X$, $x_1 \leq x_2 \leq \ldots \leq {\cal X} $, we have 
$SF({x_1}) \leq SF({x_2}) \leq  \ldots \leq SF({\cal X})$ 
and we define $x(J)$ in complete analogy with the previous paragraph. 
Then we define the {\em $x$-partial sum} of the path integral 
\be
\label{partialsum}
\rho_x(j) = 
\sum_{J: \partial J=j , x(J) \leq x} |J|_x .
\ee
Obviously for $x = {\cal X}$ we have $ \rho_x(j) = \rho_{ _{\cal X}}(j)$. Another simple property of this definition is that 
the weight of a spin foam takes the same value for any 
two 
$x, x'$ sublattices of $ {\cal X}$, 
$|J|_x = |J|_{x'} = | J |_{ _{\cal X}}$. This property of the weights 
is clear in our context. In an article by Bojowald and Perez \cite{NoAnomaly} it 
was called a ``no anomaly" requirement for spin foam models. 
However, 
it is also clear that if $\rho_{ _{\cal X}}(j)$ is not a partial sum but the partition function of an effective theory defined at ${\cal X}$ there is no reason for requiring that $|J|_{ _{\cal X}}$ be independent of ${\cal X}$. 

The use of {\em embedded lattices} is natural because Wilson loops are gauge invariant observables of the connection on which  diffeomorphisms, symmetries of general relativity, act simply by shifting the loop. This fact lead to the formalism now known as canonical loop quantization  \cite{LQreviews}. 
Here we review its formulation in terms of the nested sublattice structure (originally developed by Ashtekar and Lewandowski \cite{ALmeasure}), and propose a natural ansatz for its extension to the covariant formalism. 

Consider the space manifold to be $\Sigma$. 
We do not introduce a cut-off and will define the lattice gauge theory of ``the finest embedded lattice." 
All the lattices embedded in $\Sigma$  form a family of ``sublattices" $S(\Sigma)$ which is partially ordered by inclusion and directed in the direction of refinement. 

A connection on a lattice induces a connection on any of its sublattices. Then we have the following natural maps 
${\rm Fun}({\cal A}_{\gamma_1}) \rightarrow {\rm Fun}({\cal A}_{\gamma_2}) \rightarrow 
\ldots \rightarrow {\rm Fun}(\bar{\cal A}_{\Sigma})$ (denoted by 
$i^*_{\gamma' \gamma}$). 
The compatibility of measures 
holds again and defines the measure $d\mu_{AL}$ on $\bar{\cal A}_{\Sigma}$ \cite{ALmeasure}. This allows for the structure of nested Hilbert spaces (\ref{nestedH})
to hold and to define  ${\cal H}_{\Sigma } = 
L^2(\bar{\cal A}_{\Sigma}/ \bar{\cal G}_{\Sigma}, d\mu_{AL})$ as the Hilbert space containing all the ${\cal H}_{\gamma}$ for any embedded lattice $\gamma \subset \Sigma$. 
There is a very useful description of this space in terms of the spin network basis 
\[
{\cal H}_{\Sigma} = 
\hbox{\rm co-}\hspace{-.17cm} \lim_{\gamma \to \Sigma} {\cal H}_{\gamma} = 
\hbox{\rm co-}\hspace{-.17cm} \lim_{\gamma \to \Sigma} 
\C [ \{ j \in SN(\Sigma)| \gamma(j) \leq \gamma \}] = 
\C [SN(\Sigma)] , 
\]
where $SN(\Sigma)$ is defined as the set at the end of the infinite tower of inclusions $SN({\gamma_1}) \leq SN({\gamma_2}) \leq  \ldots \leq 
SN({\gamma_n}) \leq  \ldots \leq SN(\Sigma)$ for any sequence of nested embedded lattices $\gamma_1 \leq \gamma_2 \leq  \ldots \leq 
\gamma_n \leq  \ldots$. 
Equivalently, $SN(\Sigma)= \cup_{\gamma} SN({\gamma}))$ where the union runs over the set of lattices embedded in $\Sigma$. 

Connection and momentum operators are defined by the strategy sketched previously for nested sublattices. 
This completes the 
canonical kinematics of loop quantized theories. 
It achieves its motivating goal:  
its Hilbert space ${\cal H}_{\Sigma }$ 
does not involve the choice of a
discretization that would add a foreign background structure, and it hosts a 
unitary representation of ${\rm Diff}(\Sigma)$. 

Now we propose a covariant extension to loop quantization. The resulting models will also be called {\em embedded spin foam models}. 

Consider a spacetime manifold with boundary $M$. 
The set of embedded spin foams $SF(M)$ is defined in complete analogy with the set of embedded spin networks; that is, 
$SF(M)= \cup_x SF(x)$ with $x$ an embedded spacetime lattice in $M$%
\footnote{
A notion of embedded spin foam equivalent to ours is included in the general definition of {\em spin foam} by Baez \cite{SF}. We use a definition based on the sublattice structure because it suggests the ``summation prescription" (\ref{sprescription}) for the partition function. 
}. 
We assume that lattice gauge theory gives us the spin foam weights $|J|_M$ from the UV-limit of a collection of effective theories. We do not know how to construct these weights, we assume that they are given to us. 
Even if we work in the $q$-deformed theory (with $q$ a root of unity) with the aim
of regularization, $SN(\partial M)$, $SF(M)$ 
and $SF(M, j)= \{ J \in SF(M) | \partial J = j \}$ 
are uncountable infinite sets. Then 
it is not trivial to make sense of the spin foam expansion of the path integral. 
To give an answer we note that for any
sublattice $x \in S(M)$ 
$\rho^n_x (j)= \rho_x(j) / \! \rho_x(j=0)$ 
is a quotient of series%
\footnote{
Even if $\rho_x(j)$ diverges there could be a way to make sense of $\rho^n_x (j)$. 
}
(in the $q$-deformed case a quotient of finite sums). 
Thus, our definition of embedded spin foam models comes along with the following ``summation prescription" 
\be
\label{sprescription}
\rho^n_M (j)= \lim_{x \to M} \rho^n_x (j) , 
\ee
meaning that such a limit exists and equals $\rho^n_M (j)$ if for any $\epsilon >0$ there is a big enough $x \in S(M)$ such that for any $x' \geq x$ we have 
$| \rho^n_{x'} (j) - \rho^n_M (j) | < \epsilon $. 

Again we remark that the weights $|J|_x = |J|_{x'} = |J|_M$ do not depend on the 
sublattice $x \in S(M)$ defining the partial sum $\rho_x (j)$; 
this property was called 
the ``no anomaly" requirement by Bojowald and Perez (for a discussion see the paragraph following equation (\ref{partialsum})). 
Note that if the above limit exists the space of functions of the boundary data is 
$\C [SN(\partial M)] = {\cal H}_{\partial M}$, the kinematical Hilbert space of canonical loop quantization. 

For topological
theories it happens that there is a family of sublattices with the property that 
$\rho^n_x$ is an exact effective theory for $\rho^n_M$. 
Since this family of sublattices has the property that once $x$ is in the family any $x' \geq x$ also belongs to the family, the limit in equation (\ref{sprescription})
stabilizes to a constant 
which implies that the theory is indeed defined by the 
summation prescription (\ref{sprescription}) \cite{SFC}. For theories with local degrees of freedom the limit should be approached in a less trivial way. 

To conclude we give a list of straightforward consequences of this interpretation of loop quantization: 
\begin{enumerate}
\item  A loop quantized theory is a theory defined on the finest embedded lattice. Then, from the lattice gauge theory perspective, it is an  effective theory for
phenomena happening at the finest possible length scale. In other words, the theory should be thought of as a {\em fundamental theory}. 
This interpretation is possible only if the values of the 
coupling constants $\lambda^i(M)$ defining it are those of an UV-attractive fixed point of the renormalization group flow. 

A method to
define effective theories and an implementation of the renormalization
group compatible with the kinematics of loop quantization is 
essential from this perspective. Without it loop quantization is a framework applicable to theories when the values of their coupling constants are those of a fixed point, but it lacks a method to find this fixed point. 
As mentioned earlier, loop quantization is geared towards background independent theories, but in principle it should also be able to treat theories that use a background metric. Thus, our work in defining effective theories and a renormalization group flow is that of 
extending well known notions of lattice gauge theory to make them applicable for background free theories. This task involves a major reformulation since the very notion of scale needs to be extended to include background independent theories. 
Work in this direction is in progress  \cite{decimation}. (For related work on the renormalization of spin foam models see 
\cite{renormSF}.) 

\item  
In lattice gauge theory the canonical and covariant frameworks are compatible. 
Once we are working with a single lattice  
the space of boundary data of the effective path integral is identified with the kinematical Hilbert space of the effective canonical lattice theory. 
At every scale we have a unified understanding of the kinematics of the canonical and covariant lattice formulations. 
In this letter we state that 
if the renormalization group flow defines a fundamental theory 
this relationship is also present in the theory at the finest lattice/``smallest length scale." This fundamental theory is described in the canonical side by a loop quantized theory and in the covariant side by an {\em embedded} spin foam model. 

\item  If the spin foams of a spin foam model are to be viewed as embedded in spacetime, the weight system needs to obey the tight restrictions 
$|J|_x = |J|_{x'} = |J|_M$ 
for any $x, x' \geq x(J)$, which were 
previously called ``no anomaly" conditions \cite{NoAnomaly}. 
{\em However, if one is talking about an effective spin foam model based on a discretization $\cal X$, the weight $| J |_{ _{\cal X}}$ would be $\cal X$ dependent.} 

\item  If an operator is defined following the usual canonical loop quantization techniques, the resulting operator acts naturally on ${\cal H}_{\Sigma}$. 
The operator can also be defined specifying its action on each ${\cal H}_{\gamma}$, but all these actions must be compatible because these Hilbert spaces share states. 

In the context of spin foam models there is a proposal for geometric operators (area operator in four dimensions and length operator in three dimensions) \cite{0isthere} whose action on a spin network of $j \in {\cal H}_{\gamma}$ 
has non zero contributions from edges colored with spin zero (proportional to 
$ j + 1 / 2$). Clearly the action of these operators on $j$ depends on whether it is seen as an element of ${\cal H}_{\gamma}$ or of 
${\cal H}_{\gamma(j)}$.  
These operators might be correct in an effective theory living on $\gamma$, but 
they can not be used directly to define an operator of the ``fundamental" theory on ${\cal H}_{\Sigma}$.  
\end{enumerate}

We acknowledge inspiring discussions with Elisa Manrique, Robert Oeckl and Axel Weber. This work was partially supported by CONACyT grant 40035-F and DGAPA grant IN108803-3. 


\end{document}